\documentclass[conference]{IEEEtran}
\usepackage{amsmath}
\usepackage{graphicx}
\usepackage{booktabs}
\usepackage{multirow}
\usepackage{stfloats}
\usepackage{xcolor}

\definecolor{Paired-2}{RGB}{166,206,227}
\definecolor{Paired-1}{RGB}{31,120,180}
\definecolor{Paired-4}{RGB}{178,223,138}
\definecolor{Paired-3}{RGB}{51,160,44}
\definecolor{Paired-6}{RGB}{251,154,153}
\definecolor{Paired-5}{RGB}{227,26,28}
\definecolor{Paired-8}{RGB}{253,191,111}
\definecolor{Paired-7}{RGB}{255,127,0}
\definecolor{Paired-10}{RGB}{202,178,214}
\definecolor{Paired-9}{RGB}{106,61,154}
\definecolor{Paired-12}{RGB}{255,255,153}
\definecolor{Paired-11}{RGB}{177,89,40}

\begin{document}

\title{Low-Latency Software Polar Encoders and Decoders for Short Blocklengths}

\author{\IEEEauthorblockN{Mathieu Léonardon\IEEEauthorrefmark{1},
         Mohammed El Houcine Ayoubi\IEEEauthorrefmark{1},
         Adrien Cassagne\IEEEauthorrefmark{2},
         Romain Tajan\IEEEauthorrefmark{3} and
         Camille Leroux\IEEEauthorrefmark{3}
}
\IEEEauthorblockA{\IEEEauthorrefmark{1}IMT Atlantique, Lab-STICC, UMR CNRS 6285,
  29238 Brest, France}
\IEEEauthorblockA{\IEEEauthorrefmark{2}LIP6, Sorbonne Université, CNRS, UMR 7606
  Paris, France}
\IEEEauthorblockA{\IEEEauthorrefmark{3}University of Bordeaux, Bordeaux INP, IMS
  Laboratory, UMR CNRS 5218, Bordeaux, France}}

\maketitle

\begin{abstract}
This paper presents our low-latency Polar code encoders and decoders developed
for the 2025 International Symposium on Topics in Coding (ISTC 2025) contest,
which challenges participants to implement the fastest possible channel code
encoders and decoders in terms of average and maximum latency on a CPU target.
Our solution is based on Polar codes with an Adaptive Successive Cancellation
List (ASCL) decoder.

We introduce a novel ASCL unrolled decoder generator. We conduct an extensive
exploration of the design space, including code construction, CRC selection, and
list size, to identify optimal trade-offs between signal-to-noise ratio and
decoding time across various operating points. The considered operating points
are frame error rates of \(10^{-3}\) and \(10^{-5}\), information bit lengths of
64, 128, 256, and 512, and code rates of \(1/4\), \(1/2\), and \(4/5\). We also
propose an optimized bit-packed encoder.

All implementations of the encoders and decoders, along with the code
construction and the unrolled decoders generator, are released as open source in
the AFF3CT toolbox.
\end{abstract}

\section{Introduction \& Related-Works}

The ISTC 2025 Challenge advances the emerging field of software-defined channel
decoding by tasking participants with developing low-latency, high-performance
C++ encoders and decoders for short blocklength codes on single-core CPUs,
evaluated across various code rates and block sizes.

In our submission\footnote{Contest code and materials:
https://github.com/aff3ct/istc25\_contest.}, we focus on Polar
codes~\cite{Arikan2009} concatenated with a CRC. For decoding, we employ an
Adaptive Successive Cancellation List (ASCL) decoder, which first attempts
low-latency SC decoding and only falls back to SCL decoding if
necessary~\cite{Tal2011,Li2012}. This approach achieves the error-correction
performance of SCL decoding while significantly reducing average latency.

While Polar decoder optimization typically receives the most attention, we also
propose a bit-packed encoder in Section~\ref{encoder}. On the decoder side, we
leverage and refine FAST-SSC-List decoding, unrolling, and vectorization
techniques (Section~\ref{unrolled}) to reduce latency. These optimizations
enable high-throughput software decoding on general-purpose CPUs. A
comprehensive design space exploration in Section~\ref{search} identifies
optimal SNR/latency trade-offs. All source code and tools are released in the
open source AFF3CT toolbox~\cite{Cassagne2019a}.

\section{Bit-packed Encoder}
\label{encoder}

The encoder implementation relies on bit-packing and proceeds in four stages.
First, the input vector $U_N$, containing both frozen and information bits, is
generated. This is done by statically identifying constituent nodes at compile
time: rate-0 nodes are skipped, and large rate-1 nodes can be vectorized during
memory copy. Second, $U_N$ is bit-packed using SIMD instructions, specifically
\texttt{\_mm256\_cmpgt\_epi8} followed by \texttt{\_mm256\_movemask\_epi8},
which extracts the MSBs of each 8-bit element to form packed output. Third,
systematic Polar encoding is performed using vectorized XOR and shift
operations, applied intra- or inter-register depending on the node position in
the encoding graph. Finally, unpacking is achieved using a 256-entry lookup
table of 256-bit values.

\begin{figure*}[t]
  \centering
  \includegraphics[width=0.85\linewidth]{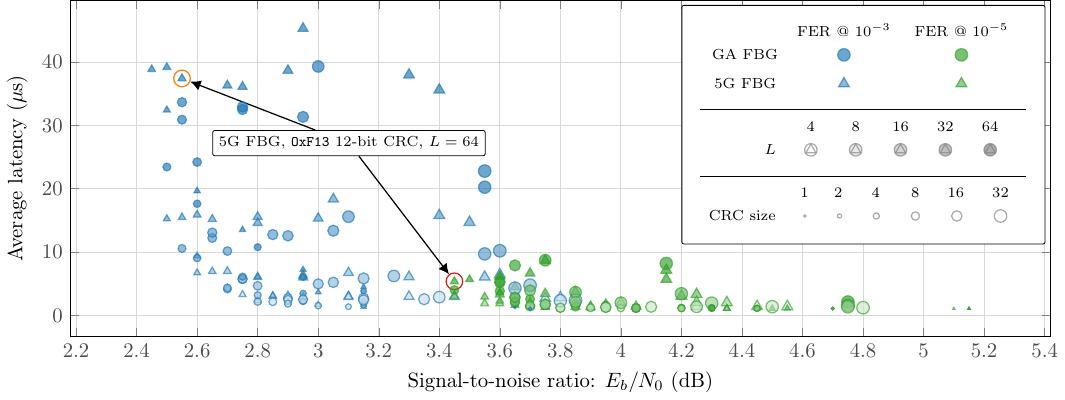}
  \caption{Space exploration of the ASCL decoder performance for $K=128$ and
    $R = 1/2$. Measured latencies from the AFF3CT generic ASCL decoder running
    on one core of an Intel Xeon Gold 6140 CPU. The selected code is highlighted
    in \textcolor{Paired-7}{orange} for FER @ $10^{-3}$ and in
    \textcolor{Paired-5}{red} for FER @ $10^{-5}$.}
  \label{fig:spider_algs}
\end{figure*}

\begin{table*}[h]
\centering
\caption{Proposed codes for the contest with the corresponding achieved
  performance (fully unrolled ASCL decoders).}
\label{tab:stats}
\resizebox{1.55\columnwidth}{!}{
{\footnotesize
\begin{tabular}{c r r l c c r r r r r}
\toprule
& & & & & & \multicolumn{5}{c}{Decoder} \\ \cmidrule(lr){7-11}
& & \multicolumn{3}{c}{Code specifications} & \multicolumn{1}{c}{Encoder} & & \multicolumn{2}{c}{FER @ $10^{-3}$} & \multicolumn{2}{c}{FER @ $10^{-5}$} \\ \cmidrule(lr){3-5} \cmidrule(lr){6-6} \cmidrule(lr){8-9} \cmidrule(lr){10-11}
Rate & $K$ & CRC size & CRC poly. & Fb. gen. & $\mathcal{L}_\text{worst}$ ($\mu$s) & $L$ & $E_b/N_0$ & $\mathcal{L}_\text{avg}$ ($\mu$s) & $E_b/N_0$ & $\mathcal{L}_\text{avg}$ ($\mu$s) \\
\midrule
\multirow{4}{*}{$1/4$} &  64 & 11 & \texttt{0x385 } & 5G & 0.20 & 64 &  2.15 dB &  2.53 &  3.35 dB & 0.55 \\
                       & 128 & 12 & \texttt{0xF13 } & 5G & 0.28 & 64 &  1.60 dB &  6.44 &  2.60 dB & 1.07 \\
                       & 256 & 12 & \texttt{0xF13 } & 5G & 0.60 & 64 &  1.25 dB & 10.58 &  2.25 dB & 1.66 \\
                       & 512 & 12 & \texttt{0xF13 } & GA & 1.01 & 32 &  1.10 dB & 14.92 &  1.85 dB & 3.53 \\
\addlinespace
\multirow{4}{*}{$1/2$} &  64 &  7 & \texttt{0x65  } & 5G & 0.14 & 32 &  3.15 dB &  0.70 &  4.60 dB & 0.33 \\
                       & 128 & 12 & \texttt{0xF13 } & 5G & 0.21 & 64 &  2.55 dB &  3.31 &  3.45 dB & 0.69 \\
                       & 256 & 12 & \texttt{0xF13 } & 5G & 0.34 & 64 &  2.15 dB &  6.55 &  3.10 dB & 1.18 \\
                       & 512 & 16 & \texttt{0x8005} & 5G & 0.68 & 64 &  1.90 dB & 11.95 &  2.70 dB & 2.12 \\
\addlinespace
\multirow{4}{*}{$4/5$} &  64 &  8 & \texttt{0x9B  } & GA & 0.15 & 32 &  5.05 dB &  0.48 &  6.45 dB & 0.16 \\
                       & 128 & 10 & \texttt{0x3D9 } & 5G & 0.20 & 32 &  4.40 dB &  1.21 &  5.35 dB & 0.41 \\
                       & 256 & 10 & \texttt{0x3D9 } & 5G & 0.34 & 32 &  4.10 dB &  2.20 &  5.00 dB & 0.67 \\
                       & 512 & 12 & \texttt{0xF13 } & 5G & 0.66 & 32 &  3.75 dB &  4.92 &  4.55 dB & 1.36 \\

\bottomrule
\end{tabular}
}}
\end{table*}

\section{Unrolled ASCL Decoder}
\label{unrolled}

The decoder implemented in this work follows the ASCL decoding principle. This
approach aims to reduce the average latency of Polar decoders without
compromising error-correction performance. Decoding begins with a fast SC
decoder. If the decoded codeword passes the CRC check, it is accepted
immediately. Otherwise, the decoder falls back to more powerful SCL decoders
with increasing list sizes $L$. This adaptive strategy significantly lowers the
average decoding time in high-SNR regimes, where SC decoding is often
sufficient, while preserving the robustness of SCL decoding in more challenging
conditions. The ASCL mechanism is well suited for short blocklength codes, where
latency is critical.

To achieve low latency, we adopt a fully unrolled decoder generation strategy.
For each Polar code configuration used in the contest, a dedicated and
specialized C++ source file is generated. This eliminates control-flow overhead
such as loops and recursive function calls, making the approach particularly
effective for short blocklengths where latency is critical.
We provide an open source decoder generator tool\footnote{Polar decoder
generator: https://github.com/aff3ct/polar\_decoder\_gen}, which supports SC,
SCL, and ASCL decoding modes. The tool produces optimized C++ code tailored to
each frozen set, enabling reproducibility and integration into larger systems.

\section{Search space and exploration}
\label{search}

To identify optimal trade-offs for the ISTC 2025 contest, we explored various
code and decoder configurations, including different code constructions (GA,
5G)~\cite{Trifonov2012,3GPP38212}, CRC lengths and polynomials -- including some
polynomials that do not correspond to standard CRCs -- and list sizes $L$. This
search helped uncover ``sweet spots'' balancing latency and error-correction
performance. An example of the results obtained for \(K = 128\) and \(N = 256\)
is shown in Figure~\ref{fig:spider_algs}.

\section{Performance results}

\paragraph{Testbed}

A Minisforum AtomMan X7 Ti PC (Intel Ultra 9 185H, 16 x86-64 cores with
6~p-cores @ 5.1 GHz (one core used), 8~e-cores @ 3.8 GHz (unused), and
2~LPe-cores (unused), 32 GB DDR5 @ 5600 MT/s) running Linux (Ubuntu 25.04
kernel~6.14, g++ 14.2.0) has been used.

\paragraph{Results}

Table~\ref{tab:stats} summarizes the proposed Polar code configurations
submitted for the contest, along with their measured performance metrics. The
table provides details on the code rate, information block length \(K\), CRC
size and polynomial, code construction method, worst-case encoder latency, list
size \(L\), and decoder performance in terms of Frame Error Rate (FER) at
\(10^{-3}\) and \(10^{-5}\) thresholds, together with corresponding average
decoding latencies.

\section{Conclusion}

This work was conducted in the context of the ISTC 2025 low-latency software
coding challenge, targeting high-performance implementations of channel encoders
and decoders for short blocklengths on general-purpose CPUs. We presented a
complete open source toolchain based on Polar codes, including a bit-packed
encoder and a fully unrolled ASCL decoder generator. Design space exploration
revealed competitive configurations balancing latency and error performance. All
implementations are available within the AFF3CT framework to foster
reproducibility and further research.

\bibliographystyle{IEEEtran}
\bibliography{article}

\end{document}